\documentclass[runningheads]{llncs}
\usepackage[T1]{fontenc}
\usepackage{graphicx}
\usepackage{xcolor}
\usepackage[acronym,nonumberlist]{glossaries}
\usepackage{array}
\newcolumntype{L}[1]{>{\raggedright\arraybackslash}p{#1}}

\usepackage{amssymb}

\makeglossaries 
\newglossaryentry{component}
{
    name=component,
    description={a part or element of a larger whole, especially part of a machine or vehicle}
}
\newglossaryentry{agent}
{
    name=agent,
    description={an entity being trained}
}
\newglossaryentry{environment}
{
    name=environment,
    description={the surrounding or conditions in which an entity operates}
}
\newglossaryentry{adversary}
{
    name=adversary,
    description={one's opponent in a contest, conflict, or dispute}
}
\newglossaryentry{entity}
{
    name=entity,
    description={a thing with distinct and independent existence}
}
\newacronym{a2c}{A2C}{Advantage Actor Critic}
\newacronym{a3c}{A3C}{Asynchronous Advantage Actor Critic}
\newacronym{ai}{AI}{Artificial Intelligence}
\newacronym{ann}{ANN}{Artificial Neural Network}
\newacronym{cli}{CLI}{Command Line Interface}
\newacronym{coa}{CoA}{Course of action}
\newacronym{cps}{CPS}{Cyber Physical System}
\newacronym{drl}{DRL}{Deep Reinforcement Learning}
\newacronym{dqn}{DQN}{Deep Q-Network}
\newacronym{ffi}{FFI}{Norwegian Defence Research Establishment}
\newacronym{hat}{HAT}{Human-autonomy teaming}
\newacronym{ids}{IDS}{Intrusion Detection System}
\newacronym{ips}{IPS}{Intrusion Prevention System}
\newacronym{irs}{IRS}{Intrusion Response System}
\newacronym{ml}{ML}{Machine Learning}
\newacronym{ooda}{OODA}{Observe-orient-decide-act}
\newacronym{ppo}{PPO}{Proximal Policy Optimization}
\newacronym{rg}{RG}{Reality Gap}
\newacronym{rl}{RL}{Reinforcement Learning}
\newacronym{ros}{ROS}{Robot Operating System}
\newacronym{sb3}{SB3}{Stable Baselines3}
\newacronym{s2s}{S2S}{Sim-to-Sim}
\newacronym{s2r}{S2R}{Sim-to-Real}
\newacronym{td}{TD}{Temporal Difference}
\newacronym{ugv}{UGV}{Unmanned Ground Vehicle}
\newacronym{vsc}{VSC}{Visual Studio Code}
\begin{document}
\title{Exploring reinforcement learning for incident response in autonomous military vehicles}

\author{Henrik Madsen\inst{1,2} \and Gudmund Grov\inst{1,2}\orcidID{0000-0001-8837-5496} \and Federico Mancini\inst{1}\orcidID{0009-0003-4680-344X} \and Magnus Baksaas\inst{1} \and Åvald Åslaugson Sommervoll\inst{1,2}}

\institute{
Norwegian Defence Research Establishment (FFI), Kjeller, Norway\\
\email{\{Gudmund.Grov,Federico.Mancini,Avald-Aslaugson.Sommervoll,Magnus.Baksaas\}@ffi.no}\\
\and
University of Oslo, Norway
}

\titlerunning{Exploring RL for incident response in autonomous military vehicles}
%
%
\authorrunning{Madsen, Grov, Mancini, Baksaas and Sommervoll}

%
%
\maketitle              
\begin{abstract}
Unmanned vehicles able to conduct advanced operations without human intervention are being developed at a fast pace for many purposes. Not surprisingly, they are also expected to significantly change how military operations can be conducted. To leverage the potential of this new technology in a physically and logically contested environment, security risks are to be assessed and managed accordingly. Research on this topic points to autonomous cyber defence as one of the capabilities that may be needed to accelerate the adoption of these vehicles for military purposes. Here, we pursue this line of investigation by exploring reinforcement learning to train an agent that can autonomously respond to cyber attacks on unmanned vehicles in the context of a military operation. We first developed a simple simulation environment to quickly prototype and test some proof-of-concept agents for an initial evaluation. This agent was then applied to a more realistic simulation environment and finally deployed on an actual unmanned ground vehicle for even more realism. A key contribution of our work is demonstrating that reinforcement learning is a viable approach to train an agent that can be used for autonomous cyber defence on a real unmanned ground vehicle, even when trained in a simple simulation environment.
\keywords{Autonomous cyber defence  \and Unmanned vehicles \and Reinforcement learning.}
\end{abstract}

\section{Introduction}

Drones for surveillance and targeting, underwater vehicles for mine hunting, and ground vehicles for logistics are all examples of how unmanned autonomous vehicles can support military missions. The typical advantages are the ability to collect and process large sets of sensor data faster, with higher precision, and automating dangerous or repetitive tasks while freeing up personnel. On the other hand, these vehicles' technology and their use in military operations can introduce new security risks that we currently do not know how to manage effectively. Since failing to implement adequate security may threaten both the operation they are supposed to support, and other military assets, such as classified data and technology, new security capabilities are needed to facilitate the adoption of these vehicles in a military setting.

One security capability deemed necessary to achieve a sufficient level of trust in vehicles with a higher degree of autonomy is \textit{autonomous cyber defence }\cite{NATO}. This is the ability of the vehicle itself to detect and respond to cyber attacks. The reason is that highly autonomous vehicles are likely to be used for operations where a link to a remote operator is either undesirable due to operational requirements like covertness, or unavailable due to limited network access. This would, in turn, also prevent a security operation centre (SoC) from remotely monitoring the systems' status and detecting and managing potential cyber attacks. The risk of employing a vehicle in this context without the capability of proactively protecting itself to some degree, could outweigh the operational advantages in many cases. Well-established preventive security measures -- such as encryption, integrity verification and anti-tampering -- are still needed, but are neither sufficient nor as effective on unmanned vehicles as they are for other systems, which can be kept in controlled facilities or under constant monitoring. Therefore, we explore the possibility of complementing preventive security measures with an autonomous agent that can respond to cyber security incidents, given that a reliable detection capability is in place.

While autonomous agents for cyber defence is not a new research area, most of what is available in the literature pertains to classical computer networks or internet-based threats. Some complicating factors for unmanned vehicles in military operations make developing autonomous agents also more challenging than in other contexts, as they would have to operate at the intersection between cyber security, safety and mission assurance. Anomalies in any of those domains could be a symptom of a cyber attack, and responses aimed at handling an incident in one of the domains, could negatively affect the others. For instance, a light-diode blinking in an anomalous manner could indicate an attempt to exfiltrate data by using it as a side channel. However, shutting down the computing unit controlling the diode could cause other actuators to stop working properly, and causing the vehicle to crash. Hence, response actions need to be assessed based on many conflicting concerns.
Reinforcement learning seems like a promising approach to create such an autonomous agent that can consider these different conflicting concerns. Our overall research hypothesis is:
\begin{quote}
    \textit{Reinforcement learning can produce autonomous agents that can respond to cyber incidents while balancing conflicting security concerns that autonomous vehicles may have to consider in a military operation.}
\end{quote}
In this paper, we take some initial steps to address this hypothesis.
We focus on incident response and a limitation of our work is that we assume that detection is achieved by other means. The contributions of this paper are that we develop, apply and compare different reinforcement learning agents for an Unmanned Ground Vehicle (UGV) for different simulation environments as well as on a real military UGV. The applied nature of work, where the agents are used on a real UGV, should be seen as our key contribution. The paper is based on a recently published master thesis by the first author \cite{MadsenMsc}.

The following section introduces reinforcement learning and surveys the existing use of reinforcement learning for incident response. Section \ref{sec:problem} details our operational context, the UGV and the experimental design and implementation. The results are given in section \ref{sec:discussion}, before we conclude the paper in 
section \ref{sec:conclusion}.

\section{Reinforcement learning for incident response}
\label{sec:reinforcement}


Reinforcement learning (RL) is a machine learning (ML) paradigm that learns what to do, how to map situations to actions, in order to maximize a numerical reward signal \cite{sutton2018reinforcement}. Here, an \gls{agent} is not told what actions it should take, but instead explores an \gls{environment} by selecting actions and observing the change. This observed change can, and often will, result in a change of state. The agent's state gives how the agent percieves the environment at a given time-step. The RL problem is often seen as an agent that interacts with the environment through actions, and the environment responds with a new state and a reward for the change. This reward can be positive, giving positive reinforcement for the selected action in the given state, or it can be negative giving negative reinforcement for the action in the given state. In RL, 
a \textit{policy} defines an \gls{agent}s behaviour for a given state, i.e. how the the agent selects actions, and the goal of RL is to find a good, or ideally optimal, policy.
A \emph{reward signal} is an immediate reward, and provides the logic in which actions are good or bad in the short term.
In contrast to the reward signal, a \emph{value function} also tries to consider long-term rewards.

The \emph{exploration rate} ($\epsilon$) hyperparameter\footnote{In machine learning, a hyperparameter specifies details of the learning process, while a parameter is part of the model being learned itself.} describes how much an agent will explore an environment or exploit the knowledge it has learned about the environment \cite{medium_epsilon}.
The \emph{learning rate} ($\alpha$) hyperparameter controls the rate in which the model learns new information, 
while the \emph{discount factor} ($\gamma$) determines the importance of future rewards compared to immediate rewards. 
A challenge in RL is balancing exploration and exploitation -- to obtain most reward it needs to commit to the 
best action(s) that it has used in previous states, however before it can exploit it needs to explore the different actions and progressively favour the actions that are best \cite{sutton2018reinforcement}. 
In \emph{epsilon-greedy} action selection \cite{gfg_eps}, the agent reduces the probability of exploiting over time.
We consider two  RL algorithms:
\begin{description}
    \item[Q-learning] was one of the early breakthroughs in the field of RL. The algorithm involves a table (Q-table) which contains values for each state-action pair that is provided by the environment. With more states and actions, the Q-table becomes larger, so while it is highly effective for simple reinforcement learning problems, there can be memory constraints for more complicated problems.
    \item[Deep Q-Network (DQN)] addresses some table constraints by approximating them by a deep neural network. This abstraction is used in most modern reinforcement learning algorithms as large and complex continuous state and action spaces often appear in the real world \cite{nguyen2021deep}. 
    DQN comes under the umbrella of \emph{Deep Reinforcement Learning (DRL)} as it combines RL with deep learning.   
\end{description}

 \begin{table}[htp]
    \centering
        \begin{tabular}{ L{5cm} | L{2cm} | L{5cm} }
        \hline
        \textbf{Paper} & \textbf{Year} & \textbf{Focus Area} \\
        \hline
        Ridley \cite{ridley2018machine} & 2018 & Q-learning \\
        Stefano \& Ramachandran \cite{stefanova2018off} & 2018 & Q-learning \\
        Iannucci et al. \cite{iannucci2019towards} & 2019 & Q-learning \\
        Huang et al. \cite{huang2019game} & 2019 & Q-learning \\
        Paul et al. \cite{paul2019learning} & 2019 & Q-learning \\
        Elderman et. al. \cite{elderman2017adversarial} & 2017 & DRL and Q-learning \\
        Nguyen \& Reddi \cite{nguyen2021deep} & 2021 & DRL review \\
        Nyberg et al.\cite{nyberg2022cyber} & 2022 & DRL \\
        Bashendy et al.\cite{bashendy2023intrusion} & 2023 & Survey \\
        Salvato et. al.\cite{ieeRealityGap} & 2021 & \gls{rg}\\
        \hline
        \end{tabular}
    \caption{Summary of RL literature for incident response.}
    \label{tab:papers_catalogue}
\end{table}
 
 Work on RL for incident response is in it its infancy, at least when compared with the use of ML for incident response. This is even more true for military usage or use in cyber-physical systems. Table \ref{tab:papers_catalogue} summarises the most relevant literature.  

Ridley \cite{ridley2018machine} constructed an autonomous cyber-defense system to defend and control a network. This work focused on reasoning and learning methods to train an RL agent to defend a network from being compromised by an attacker. Here, there were three states with attacking and defending actions that could select between three or two different actions, respectively.
Stefano \& Ramachandran \cite{stefanova2018off} used Q-learning to find a good response policy in network security. They used a  simple network with just two network states and two actions, and found that similar polices were learned for all their $24$ attack scenario. 
Bashendy et al.\cite{bashendy2023intrusion} have pointed out that this solution is too simple to be suitable for cyber-physical systems (CPS).
Iannucci et al. \cite{iannucci2019towards} proposed a similar approach to Stefano \& Ramachandran, where  Q-learning is used to train an agent to operate in an environment, represented as a three-tier web application, which  
Bendy et al. \cite{bashendy2023intrusion} argues achieves a near-optimal policy and adapts to the different scenarios.
Similarly to Ridley's work, Huang et al. \cite{huang2019game} used  Q-learning for choosing optimal strategies in a two-player zero-sum stochastic game model to mitigate and defend against cyber attacks in industrial CPS. While effective, Bashendy et al. has pointed out that their approach assumes that 
the attacker and defender players have complete information on each other's actions.
Paul et al. \cite{paul2019learning} use Q-learning to find good defender actions in an environment simulating a power system. They also model the problem as a two-player zero-sum repeated game. The game considers factors like attacker and defender costs, budgets, and players strengths.
Elderman et al. \cite{elderman2017adversarial} created a simulation illustrating cyber security problems as a game with an attacker and defender, where both of them would have incomplete information. In the simulation, the attack does not have knowledge of how the network is structured, but the defender has. Here, Monte Carlo, Q-learning and neural networks were explored, with  Monte Carlo (with softmax) achieving the best results. 

Traditional RL does not scale sufficiently to handle large and complex cyber security problems.
There is a large number and diversity of CPS applications  \cite{usnsf_cps}, which additional requirements for
correctness \cite{nguyen2021deep}. CPS use different sensors, cameras, etc. They are also vulnerable to cyber attacks seeking to distort and manipulate captured sensor data. 
Defending against such attacks becomes highly relevant to our problem using an UGV, as autonomous driving capabilities rely on correct input data from LiDAR and cameras. Testing such scenarios could be done by following Gupta \& Yang's \cite{gupta2018adversarial} , where they looked at ways to increase robustness of autonomous systems using adversarial examples.
Cyberspace involves many different cyber components. Different components can potentially interact with each other, and as a result create many different scenarios, thus making the observation space larger and more complex \cite{nguyen2021deep}.
Nyberg et al.\cite{nyberg2022cyber} focus on intrusion response and the decision-making of an intrusion response system (IRS). Committing to a defensive action is not free of consequences, and a wrong action can quickly result in unnecessary operational costs. They use the Meta Attack Language (MAL) to produce attack graphs to simulate the environment, with promising initial results for DRL, including DQN.

Nguyen \& Reddi \cite{nguyen2021deep} highlights that one of the main challenges when implementing RL for CPS is the scarcity of realistic simulations -- the importance of realistic simulations is also highlighted by Salvato et al. \cite{ieeRealityGap}. They also pinpoint what they call the \emph{reality gap} (RG) in using RL in 
the context of robotic control: training time in robotic environments often becomes much more significant than in simulated environments. Moreover, a live training phase on a real robot could lead to unsafe actions when exploring, and that simulators are often essential to reduce the training time. However, the cost of creating an environment as close to reality as possible is high \cite{salvato2021characterization}. We return to this aspect in section \ref{sec:approach}.

\section{Operational context \& experimental design}
\label{sec:approach}\label{sec:problem}

\begin{figure}
    \centering
    \includegraphics[width=0.4\linewidth]{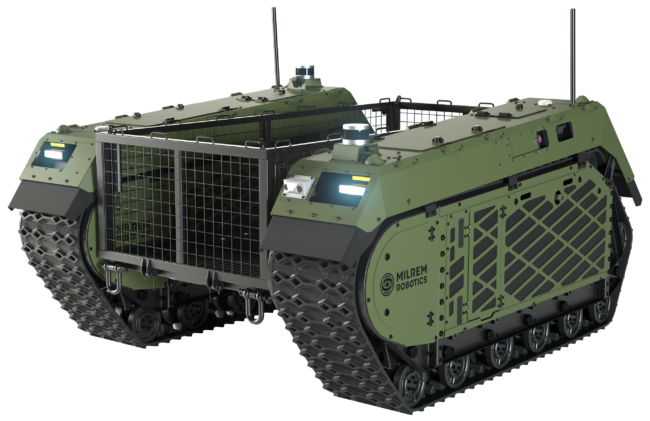}
    \vspace{-10pt}
    \caption{The Milrem THeMIS UGV \cite{themis_ugv} used for the experiments in this paper. }
    \label{fig:ugv}
\end{figure}

For our experiments, we use an actual Unmanned Ground Vehicle which is actively being deployed in Ukraine \cite{milrem_ukraine}: The Milrem THeMIS \cite{themis_ugv} (shown in Figure \ref{fig:ugv}). The Moscow-based Center for Analysis of Strategies and Technologies (CAST) announced a reward to anyone who manages to capture and deliver a Milrem THeMIS to the Russian Armed Forces \cite{ugv_reward}. This supports our motivation that such vehicles will become important intelligence targets and that collected intelligence can be used for more targeted attacks against them later, possibly as part of a (joint) operation. Such possible attacks were explored in a previous work on the same vehicle \cite{kaasen2022towards}. Here, we assume specifically that an opponent has managed to gain some privileged access to the vehicle's systems thanks to a successful supply chain compromise, so that some component will be able to initiate an attack internally without external intervention. 
The focus is thus exclusively on response with the following assumption:
\begin{quote}
    \textit{We assume that an attack is detected (by other means) and our objective is to find a suitable response action.}
\end{quote}

\begin{figure}
    \centering
    \includegraphics[width=0.4\linewidth]{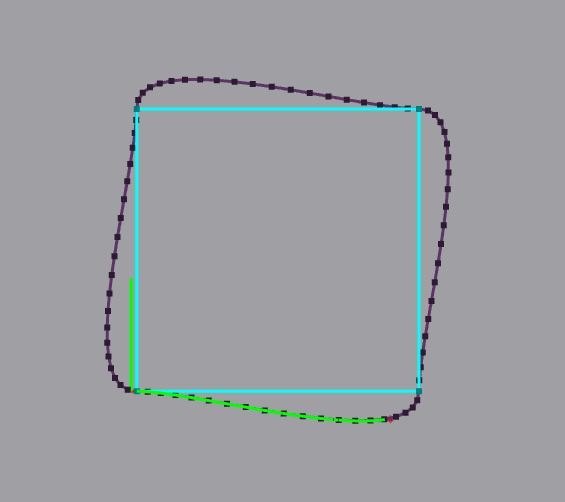}
    \includegraphics[width=0.55\linewidth]{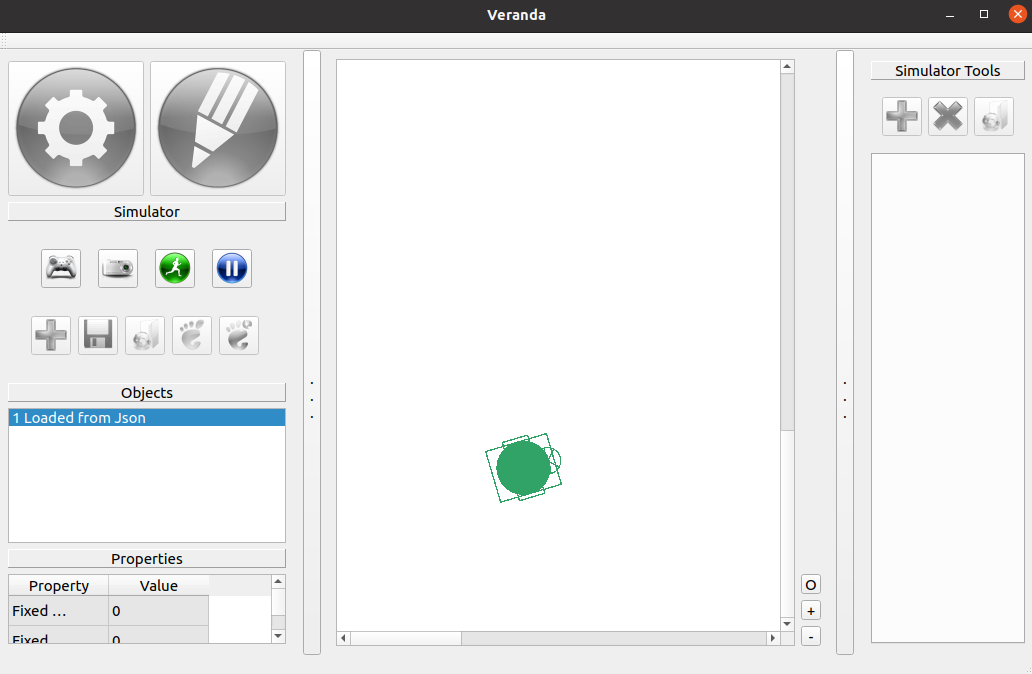}
    \caption{Simulation visualization. The left side shows our simulation using the \texttt{Mapviz} tool, while the right side shows
    our simulation with the \texttt{Veranda} tool.}
    \label{fig:simenv}
\end{figure}

This UGV is modular in functionality, which has proven to be important in reducing the risk of losing human lives. It uses natively ROS2 \cite{ros2} for internal communication between sensors, control modules and actuators, but the version used for our tests has also been extended with custom autonomy capabilities and additional navigation sensors that communicate with each other by using the ROS1 system \cite{DevelopmentOfAutonomy}. ROS systems consist of nodes that communicate through \emph{topics}. A topic could be about a component like "Brakes", and messages about the state of that component can be published and read by any other ROS node on the vehicle subscribing to that topic. Accompanying the UGV, there is a simulation tool \cite{DevelopmentOfAutonomy} and two visualization tools. The first is visualization tool is \texttt{Veranda}, which is an open-source 2D robotics visualization tool written to interface with applications via the ROS2 communication layer \cite{veranda_git}. The THeMIS Simulator publishes the individual simulated track speeds to Veranda, allowing the UGV to be simulated in a virtual 2D world. The second tool is \texttt{Mapviz}, a ROS-based visualization tool focusing on visualizing 2D data \cite{git-mapviz}. Both \texttt{Veranda} and \texttt{Mapviz} are illustrated in figure \ref{fig:simenv}.

To model realistic operations, we used actual experiences from the deployment of the THeMIS UGV in Ukraine. Seven of these vehicles were there configured for casualty evacuation and were delivered at the end of 2022. Another seven were delivered during the second quarter of 2023 with a configuration for route clearance \cite{milrem_ukraine}. The agent had therefore the following mission objective:  
\begin{quote}
    \textit{The mission objective is to ensure that the UGV safely reaches a specific point within a given time.}
\end{quote}
The agent's main goal is therefore to ensure, or at least increase the likelihood of, this mission's success by correctly responding to adversarial activities that can cause the UGV to e.g. stop, be captured, or destroy its cargo. 

To represent adversarial activity by an attacker, random events are implemented that may occur at any time while the mission is in progress. These events are created by selecting a random component and altering its state. This may result in the UGV coming to a halt. The agent must respond correctly by returning the compromised component to its original state so the UGV can drive again. The cost of response actions must be considered, as changing the state of a system component may cause physical harm to that component or other components it communicates with. Furthermore, the agent must not simply exploit specific actions to attain the most reward when no threat exists. We achieve this by having the agent remain idle or monitor the environment while no adversarial activity has been detected.\footnote{This is represented by the action called \texttt{Do nothing}, which does not change the environment, as discussed below and shown in table \ref{tab:ugv_attributes}.} The available response options may vary based on travel time constraints and how they can affect, respectively, the people being transported or the payload mounted on the UGV. 

The two RL algorithms we have explored, Q-learning and DQN, are based repsectively on Sutton \& Barto's pseudocode \cite{sutton2018reinforcement} and Stable-Baselines3 (SB3). In order to quickly prototype the agents, or, in other words, to quickly develop and implement the code for applying RL to intrusion response in a CPS, the Python programming language was used. 

\begin{figure}
    \centering
    \includegraphics[width=0.8\linewidth]{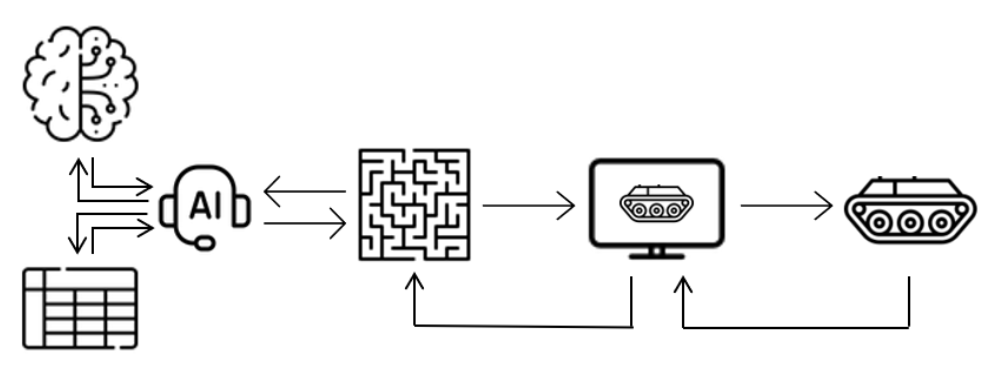}
        \vspace{-10pt}
    \caption{Overall approach. The AI agents are trained using RL and deep RL in a simple simulation environment. The trained agents are then applied in a more complex integrated simulation environment before being applied in a real environment. The feedback loops indicate how the more complex environments support the training.}
    \label{fig:approach}
\end{figure}

Figure \ref{fig:approach} shows our overall approach, where these algorithms have been used in three different environments of increased complexity and realism, with a feedback loop back to help guide development in the simpler model:

\begin{description}
    \item[A simple simulation environment.] RL, and particularly Deep RL like DQN, requires many simulations. To support initial development and necessary speed in training, a simple stand-alone simulation environment that is fully decoupled from the UGV details was developed using Gymnasium \cite{gymnasium-docs}. As shown by bi-directional arrows in figure \ref{fig:approach}, this environment was used to train the RL agent.
       
    \item[An integrated simulation environment.] 
    The simple simulation environment chooses simplicity and speed over realism. Our motivation for what we have called the \emph{integrated simulation environment} is to add realism to the simulation -- however, this is at the expense of speed. Here, we have utilised the existing simulation environment for the THeMIS, as explained above. This environment was originally developed for other purposes, so we had to implement an interface/bridge between our RL agents and the UGV simulator to allow them to communicate seamlessly in real-time. Here, both \texttt{Veranda} and \texttt{Mapviz} were used, and figure \ref{fig:simenv} shows the use of them with our RL agents. The left hand side illustrate our mission where the UGV has to follow the given path.
    

    
    \item[A real environment.] A key contribution of our work is that we applied the RL agents on an actual Milrem THeMIS like the one shown in figure \ref{fig:ugv}.   
\end{description}

Accurate and realistic simulations are known to have a significant computational cost for training  \cite{ieeRealityGap}. 
To give an example, Q-learning had a training time of roughly one to two minutes when trained for $1,000$ episodes or more in the simple simulation environment, while the DQN model would use approximately $23$ minutes for the same number of episodes. With the UGV simulation, one episode would take between one and two minutes since it operates in real-time. This would mean that the model would have to train for many hours or days, depending on the difficulty and complexity of the problem. As illustrated in figure \ref{fig:approach}, we therefore pre-trained agents in the simple environment and integrated them with the UGV simulation in the integrated environment (with a feedback loop to guide the modelling and training), effectively using what is called a \emph{sim-to-sim} (S2S) approach \cite{ieeRealityGap}. These trained models were then used on the actual Milrem Themis UGV, also with a feedback loop to adjust the models. This approach, where a trained application from a simulation environment is transferred to the actual system, is called \emph{sim-to-real} (S2R) \cite{ieeRealityGap}.

For all simulation environments, the UGV is represented as a set of components, each with a state. The RL-agents have a set of possible actions to apply to change the state, and our goal is to learn an optimal policy for which action to apply for the different states in order to complete the mission. In the implementation, these are represented using arrays of named components, component states, the UGVs states, and the different possible actions of the RL-agents, as shown in  table \ref{tab:ugv_attributes}. These representations were based on combining the actions and states of Han et al. \cite{han2018reinforcement} and Ridley \cite{ridley2018machine}, and implemented following Zennaro \& Erd{\H{o}}di \cite{zennaro2023modelling}. As complexity increases by increasing the components and available actions, we have conducted two different experiments of increasing complexity have been used:
\begin{itemize}
    \item In \emph{experiment 1}, a reduced number of components and actions were used. These are listed in the second column of table \ref{tab:ugv_attributes}.
   \item In \emph{experiment 2}, we increased the number of components and actions possible as listed in the third  column of table \ref{tab:ugv_attributes}.
\end{itemize}


\begin{table}[htp]
    \centering
        \begin{tabular}{ L{4cm} | L{4cm} | L{4cm} }
        \hline
        \textbf{Lists} & \textbf{Experiment 1} & \textbf{Experiment 2} \\
        \hline
        COMPONENTS & ["Force Brake", "Generator", "High-voltage system"] & ["Force Brake", "Generator", "High-voltage system", "Heading", "Noise", "Trajectory"]\\
        \hline
        COMPONENTS\_STATES & ["OFF", "ON"] & ["OFF", "ON"] \\
        \hline
        STATES & ["Stationary", "Driving", "Goal reached"] & ["Stationary", "Driving", "Goal reached"] \\
        \hline
        ACTIONS & ["Turn on", "Turn off", "Do nothing"] & ["Turn on", "Turn off", "Do nothing", "Publish correct Noise", "Publish correct Heading", "Publish correct Trajectory"] \\
        \hline
        \end{tabular}
    \caption{Representations of states, components and agent actions used.}
    \label{tab:ugv_attributes}
\end{table}

The simple simulation environment has three custom input parameters: \linebreak \texttt{max\_timesteps}, \texttt{goal\_step} and \texttt{attack\_prob}. The  \texttt{max\_timesteps} parameter defines the maximum number of timesteps 
If the agent doesn't achieve its goal, the episode is ended and considered a failure. The \texttt{goal\_step} parameter introduces the position that the agent must reach before \texttt{max\_timestep} has been reached. If the agent reaches the goal, the episode is ended and marked as a success. 
The \texttt{attack\_probab} parameter configures the frequency of attacks conducted by the adversary. 
Each such attack is a simple command where the attacker turns one of the three components on or off, depending on their current status. Depending on the probability that we assign to the likelihood of an attack, the \texttt{make\_attack\_list()} method creates a dictionary containing exact timesteps/positions in which the UGV is attacked. This dictionary is remade every time the environment is initialized or reset.
The environment's action and observation spaces are defined as discrete spaces. 

The agents are provided with different rewards and penalties based on their actions. While the agent ensures that the UGV is in the \texttt{Driving} state, a reward of $1$ is given. Additionally, if the UGV is \texttt{Driving}, the agent gets an additional reward of $1$ for using the action \texttt{Do nothing}. If the agent commits to any other action while \texttt{Driving}, it receives a punishment of $-10$. 
If the UGV is \texttt{Stationary}, a penalty of $-1$ is given. If the agent commits to the action \texttt{Do nothing} while \texttt{Stationary}, an additional penalty of $-1$ is given. Finally, if the UGV reaches its end goal, the state is changed to \texttt{Goal reached}, and the agent receives a reward of $50$. If the agent runs out of time (uses all the time steps without reaching the goal), the mission is terminated, and the agent gets a penalty of $-10$. 


A similar approach was taken in the integrated simulation environment, albeit with some implementation changes to better integrate with the UGV simulator. 
Here, a timer was used instead of the time steps. 
The two parameters \texttt{min\_attack\_bound} and \texttt{max\_attack\_bound} replaced \texttt{attack\_probability} to introduce attacks in specific time intervals. Additionally, a thread was used to 
keep track of the distance the UGV has left before arriving at its end goal. The UGV was then set to follow a provided trajectory to ensure it continued driving. 
The timer was used 
to experiment with adversarial activity and agent responses in real-time. The environment kept track of whether an attack was happening and behaved differently depending on this. If an attack were not happening at the current step, then the environment would calculate a new time that an attack would occur within the time interval set by \texttt{min\_attack\_bound} and \texttt{max\_attack\_bound}. When a new \texttt{attack\_time} was created, the environment would have the attacker choose a random component to attack as described above. To integrate 
a \texttt{Stationary} state with UGV simulation, a new message was sent to a specified (ROS) topic. This resulted in the vehicle stopping unless it received any new control messages. 
If the topic command was changed to follow the trajectory, the \texttt{Driving} state was simulated.
The agent's goal remains the same: if it exceeded the time limit set by \texttt{max\_time}, the mission would have ended. If the agent reached the goal position before the time limit, the mission was a success. 
The RL agent responded to the observed states by changing the components' states, thus enabling the UGV to continue driving. If the components \texttt{Generator} or \texttt{High-voltage system} were switched off, a message was sent to the specified topic telling the UGV to stop. If the component was turned on, the UGV would be set to follow its current trajectory. This logic was reversed for the \texttt{Force Brake} component. 

On the physical UGV, we used the same model as in the integrated environment to validate whether this approach would be feasible also in the physical domain. A limitation is that we did not measure to which degree the agents met their mission objective or compared the different RL approaches. The results from all experiments are detailed and discussed in the next section.

\section{Results and discussion}
\label{sec:discussion}




\paragraph{Simple simulation environment.}
In the simple environment we test three competing strategies: random action (baseline), Q-learning (epsilon-greedy), and an argmax strategy. Where the baseline strategy is a strategy where an agent selects a random action at each timestep, and serves as a sanity check of the problems complexity and that the Q-learning agents are learning something meaningful.
Initial hyperparameter testing for Q-learning tested combinations of $0.1$, $0.5$ and $0.9$ for $\alpha$, $\gamma$ and $\epsilon$, and found that $\alpha = 0.1$, $\gamma = 0.9$ and $\epsilon=0.9$ gave favorable results.
\begin{figure}[h!]
    \centering
    \includegraphics[scale=0.5]{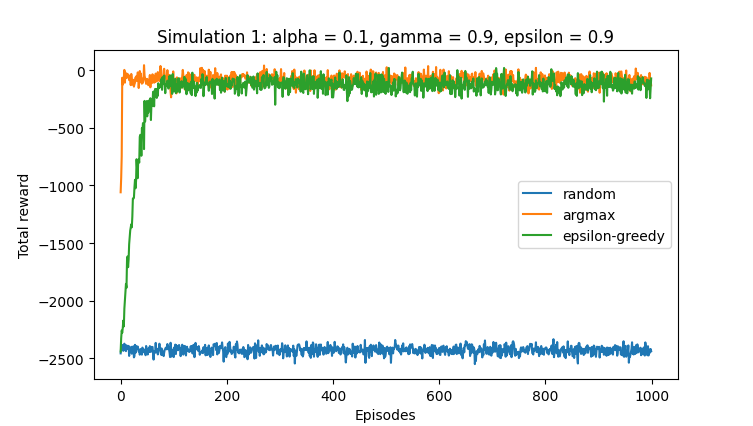}
    \caption{Experiment 1: Total reward for Q-learning for different strategies.}
    \label{fig:learning:res}
\end{figure}
The parameters for the environments was set as follows for all the experiments: \texttt{max\_timesteps} was set to $1800$,
\texttt{goal\_step} was set to $800$ and \texttt{attack\_prob} was set to $0.9$.


Figure \ref{fig:learning:res} shows the total reward for the argmax, epsilon-greedy and random choice. A key observation is that both strategies outperform random choice as was used for baseline. The result for experiment 2 is similar and not provided here. 


\begin{table}[htp]
    \centering
        \begin{tabular}{ L{3cm} | L{3cm} | L{2cm} | L{2cm} | L{2cm} }
        \hline
        \textbf{Experiment} & \textbf{Algorithm} & \textbf{Mean reward gained} & \textbf{Training time} & \textbf{Mean timesteps used}\\
        \hline
        Experiment 1 & Q-learning & -181.46 & 25.82 sec & 1610.40 \\
        Experiment 1 & DQN & -107.67 & 23.27 min & 1586.74 \\
        \hline
        Experiment 2 & Q-learning & -409.95 & 28.65 sec & 1760.32 \\
        Experiment 2 & DQN & -262.22 & 23.89 min & 1742.19 \\
        \hline
        \end{tabular}
    \caption{%
    Simple simulation environment: statistics for the RL algorithms.
    The mean reward listed it the mean reward across three episodes of the trained agents.
    The training time is the time it took to train the reinforcement learning agents.
    The mean timesteps used is the mean timesteps used across three episodes of the trained agents.
    } %
    \label{tab:tab_drl_res}
\end{table}
To extend the simulation to both experiment 1 and experiment 2 we also trained and tested a DQN agent. During testing, the parameters remained the same as above, except that \texttt{attack\_prob} was set to $0.1$, i.e. a  $10\%$ chance of being attacked.
Table \ref{tab:tab_drl_res} shows the results of the training of the reinforcement learning agents for experiment 1 and experiment 2 with the two RL algorithms.
One thing to notice is that the training of the deep learning model DQN is much slower than the more simplistic Q-learning approach. This is most likely due a difference in how they were trained. The Q-learning approach was trained for 1000 episodes with max timesteps of 1800, while the DQN was trained on  $1800 \cdot 1000 = \text{1 800 000}$ total timesteps spread across as many episodes as possible. This means that the more episodic training the Q-learning agent undertook is faster, as the agent learns to combat the adversary's attacks, leading to shorter episodes. This approach also prevents the agent getting stuck on early runs potentially wasting time. However, the fixed timestep approach that was used for DQN ensures that even though later runs are fast, it still has time to optimize its defence further. In this case, it seems like the added exploration paid off as the epsilon-greedy DQN approach both achieves a better mean reward and better mean timesteps used on the tests conducted after training. The difference, however, is not overly large and the main take-away is that both algorithms are successfull in the simple environment.

\paragraph{Integrated simulation environment.}
Both RL agents were also successful in the integrated environment. This indicates that the simple environment captured the essential elements of the more complex integrated environment. The initial promising results in the integrated environment also bodes well for the agents to also be effective in a real world setting on the UGV.


\paragraph{Experiments on the Milrem THeMIS UGV.}
The experiments on the Milrem THeMIS UGV were done over two days. The first day was mainly used for debugging, but we did already on the first day get positive results. Once the parameters and values had been adjusted correctly, the RL agents and environments were able to interact with the UGV. This could be seen from several graphical interfaces that accompanied the UGV. The data sent from the RL agents was successfully transmitted via our interface, and the states of the concerned topics were correctly changed. The transmission was also seemingly instantaneous, as the UGV switched between modes immediately. 

\begin{figure}[h!]
    \centering
    \includegraphics[width=0.7\textwidth]{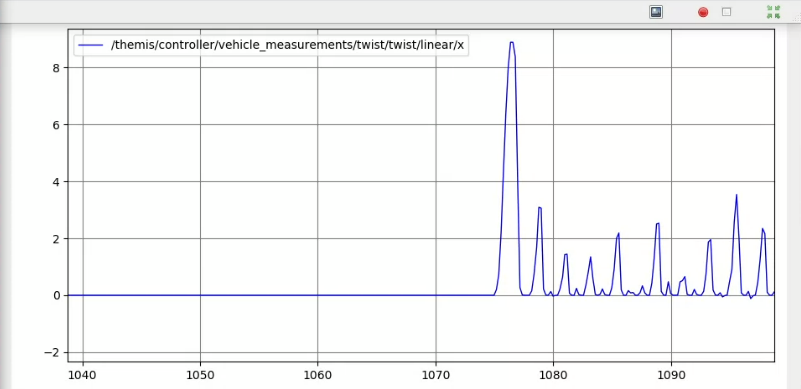}
    \caption{Velocity of UGV during evaluation on the Milrem THeMIS UGV. The graph shows how the UGV abruptly changed between accelerating and braking as a result of attacks and mitigating actions by the RL agent.}
    \label{fig:vel_tor_test}
\end{figure}

On the second day of testing, more complete tests were conducted. The initialization of agents and environments worked the same way as in the simulations, and the trajectory and commands for changing between different modes were sent correctly.
When the mission started, the UGV achieved a high velocity before being sent to a complete halt, as seen in figure \ref{fig:vel_tor_test}. Whenever the RL agents and environments switched between modes, the UGV would quickly send velocity commands to the individual tracks before stopping them whenever an attack happened. While it resulted in a rather jerky experience, with the UGV abruptly changing between accelerating and braking, it nevertheless showed that the RL worked. Some delay were observed when sending velocity commands to its tracks. The attacks and responses also happened so fast that the UGV would behave differently in real life than it did in the simulation. As a result, the UGV would remain idle for a small amount of time between the velocity commands being sent and the tracks accelerating. Still, we consider the evaluation on the Milrem THeMIS UGV successful, as it did indeed demonstrate our RL approach on a real UGV.

\paragraph{Summary \& discussion.}
 Both algorithms were successful for the two experiments and in the three environments. This emphasizes the success of the transfer learning as the agents were only trained in the simple environment and were tested successfully in the more advanced environments (S2S) and on a real (military) UGV (S2R).

\section{Conclusion \& future work}
\label{sec:conclusion}

Our overall research hypothesis for this work is that \textit{reinforcement learning can produce autonomous agents that can respond to cyber incidents while balancing conflicting security concerns that autonomous vehicles may have to consider in a military operation.}
As a target platform to address this research hypothesis, we used a Milrem Themis 4.5 UGV which has been extended with autonomous capabilities. It also comes with its own simulation tool, which has been complemented with the autonomy modules deployed on the actual UGV. To design and test S2S scenarios, we developed an environment where RL agents can be quickly prototyped, and later used in the UGV simulation tool. This allowed to later deploy the trained agents on to the actual UGV (S2R) with relative ease. We tested both traditional and deep reinforcement learning (Q-learning and DQN). The mission that was simulated was quite simple: the UGV is tasked with travelling a given route, with the main operational objective to complete it within a given time. Here, random components are made to fail with a given frequency because of cyber attacks that the agent needs to respond to by choosing from a selection of predefined actions that have effect not only in the cyber domain, but also to the physical state of the vehicle. Detection of such attacks was assumed to be in place and working as intended.

The results show that reinforcement learning gives better response outcomes than if the agent had taken random choices, albeit on simple scenarios, and thus reason to believe that this can be promising approach to produce agents for incident response on autonomous military vehicles.
The tested reinforcement learning algorithms confirm the results found in the literature, where Q-learning and DQN proved to be suitable for scenarios with discrete spaces. Most of the trained RL agents also successfully transferred for S2S and S2R scenarios, showing that it is feasible to use an operational military autonomous vehicle as a test platform already at an early prototyping stage. This will allow us to model and test more realistic threat and response scenarios that can take place in actual missions. 
We have also had to omit some details, such as additional hyper-parameter tuning and testing of other RL algorithms. 

Another valuable outcome of this work is a better understanding of what other challenges need to be resolved to give a more definitive answer to our research hypothesis. One aspect where there is room for improvement, and that may present significant challenges, is how to represent mission-critical assets in the reward function. This requires to understand the relation between mission objectives and vehicle states down to component level. This is relatively straightforward for one simple mission objective with clear dependencies, like the engine being critical to move and therefore complete the given route. If multiple components contribute to the objective in a different way under different circumstances, or multiple conflicting objectives are considered, this will likely result in much larger observation and action spaces, and could produce different results than we have seen here. If at all possible to model in a reasonable way. Another topic that should be investigated further, is how to represent components and states. A mix of strings in lists for components, and numerical values attached to them that indicated their status proved sufficient to compute a numerical representation of the overall state of the UGV to be presented to the RL agent. A more effective way could be to introduce a vector or dictionary representation of states. This could also provide better human readability and reduce computational costs, which can become significant when action and state spaces increase with more complex environments. 

To conclude, our experiments have shown promise in using reinforcement learning for autonomous incident response and given us valuable insight into what is needed to further investigate our research hypothesis. 
Promising research avenues to build on our work include handling more realistic and complex scenarios and defining reward functions that balance military operations with conflicting security concerns. The component dynamics in cyber-physical systems need to be modelled so that agents can evaluate cyber incidents and possible responses within the physical domain. Understanding the potential repercussions of applying well-established cyber security playbooks on these vehicles and the mission would also be an exciting undertaking.

\subsubsection*{Acknowledgements} This work was partly funded by the European Union as part of the European Defence Fund (EDF) project AInception (GA No. 101103385). Views and opinions expressed are however those of the authors only and do not necessarily reflect those of the European Union (EU). The EU cannot be held responsible for them.
%
%
%
\bibliographystyle{splncs04}
\bibliography{references}

\begin{thebibliography}{10}
\providecommand{\url}[1]{\texttt{#1}}
\providecommand{\urlprefix}{URL }
\providecommand{\doi}[1]{https://doi.org/#1}

\bibitem{bashendy2023intrusion}
Bashendy, M., Tantawy, A., Erradi, A.: Intrusion response systems for cyber-physical systems: A comprehensive survey. Computers \& Security  \textbf{124},  102984 (2023)

\bibitem{elderman2017adversarial}
Elderman, R., Pater, L.J., Thie, A.S., Drugan, M.M., Wiering, M.A.: Adversarial reinforcement learning in a cyber security simulation. In: ICAART (2). pp. 559--566 (2017)

\bibitem{gymnasium-docs}
Foundation, F.: Gymnasium - an api standard for reinforcement learning with a diverse collection of reference environments (2024), \url{https://gymnasium.farama.org/}

\bibitem{usnsf_cps}
Foundation, U.N.S.: Cyber-physical systems (cps) (2024), \url{https://new.nsf.gov/funding/opportunities/cyber-physical-systems-cps}

\bibitem{gfg_eps}
for Geeks, G.: Epsilon-greedy algorithm in reinforcement learning (2023), \url{https://www.geeksforgeeks.org/epsilon-greedy-algorithm-in-reinforcement-learning/}

\bibitem{gupta2018adversarial}
Gupta, A., Yang, Z.: Adversarial reinforcement learning for observer design in autonomous systems under cyber attacks. arXiv preprint arXiv:1809.06784  (2018)

\bibitem{han2018reinforcement}
Han, Y., Rubinstein, B.I., Abraham, T., Alpcan, T., De~Vel, O., Erfani, S., Hubczenko, D., Leckie, C., Montague, P.: Reinforcement learning for autonomous defence in software-defined networking. In: Decision and Game Theory for Security: 9th International Conference, GameSec 2018, Seattle, WA, USA, October 29--31, 2018, Proceedings 9. pp. 145--165. Springer (2018)

\bibitem{huang2019game}
Huang, K., Zhou, C., Qin, Y., Tu, W.: A game-theoretic approach to cross-layer security decision-making in industrial cyber-physical systems. IEEE Transactions on Industrial Electronics  \textbf{67}(3),  2371--2379 (2019)

\bibitem{iannucci2019towards}
Iannucci, S., Montemaggio, A., Williams, B.: Towards self-defense of non-stationary systems. In: 2019 International Conference on Computing, Networking and Communications (ICNC). pp. 250--254. IEEE (2019)

\bibitem{kaasen2022towards}
Kaasen, A.D., Grov, G., Mancini, F., Baksaas, M.: Towards data-driven autonomous cyber defence for military unmanned vehicles-threats \& attacks. In: MILCOM 2022-2022 IEEE Military Communications Conference (MILCOM). pp. 861--866. IEEE (2022)

\bibitem{ros2}
Macenski, S., Foote, T., Gerkey, B., Lalancette, C., Woodall, W.: Robot operating system 2: Design, architecture, and uses in the wild. Science Robotics  \textbf{7}(66),  eabm6074 (2022). \doi{10.1126/scirobotics.abm6074}, \url{https://www.science.org/doi/abs/10.1126/scirobotics.abm6074}

\bibitem{MadsenMsc}
Madsen, H.: Applying reinforcement learning for incident response in a military autonomous vehicle (2024), \url{https://www.duo.uio.no/handle/10852/112781}.

\bibitem{NATO}
Mancini, F., et~al.: Securing unmanned and autonomous vehicles for mission assurance. Tech. rep., {NATO STO T}echnical report {TR-IST-164} (NATO UNCLASSIFIED) (2022)

\bibitem{medium_epsilon}
Nair, S.: Rl series\#3: To explore or not to explore, that is the question (2020), \url{https://medium.com/analytics-vidhya/rl-series-3-to-explore-or-not-to-explore-1ff88e4bf5af}

\bibitem{nguyen2021deep}
Nguyen, T.T., Reddi, V.J.: Deep reinforcement learning for cyber security. IEEE Transactions on Neural Networks and Learning Systems  (2021)

\bibitem{nyberg2022cyber}
Nyberg, J., Johnson, P., M{\'e}hes, A.: Cyber threat response using reinforcement learning in graph-based attack simulations. In: NOMS 2022-2022 IEEE/IFIP Network Operations and Management Symposium. pp.~1--4. IEEE (2022)

\bibitem{paul2019learning}
Paul, S., Ni, Z., Mu, C.: A learning-based solution for an adversarial repeated game in cyber--physical power systems. IEEE Transactions on Neural Networks and Learning Systems  \textbf{31}(11),  4512--4523 (2019)

\bibitem{git-mapviz}
Reed, P.J., Faconti, D., Venator, E., Anthony, D., Alton, N., Matthew, jgassaway, Towler, J., Vincent, R., Shapkin, A., Alban, M., Arguedas, M., Goertz, A., jbdaniel18, Holt, B., agyoungs, mattrich37, Snider, D., Grogan, M., DangitBen, Brothers, R., austindodson, Logan, S.K., jreyes512, Kleiser, D., jhdcs, Deric, A., camjaws, D'Souza, D.: Mapviz (2024), \url{https://github.com/swri-robotics/mapviz}

\bibitem{ridley2018machine}
Ridley, A.: Machine learning for autonomous cyber defense. The Next Wave  \textbf{22}(1),  7--14 (2018)

\bibitem{milrem_ukraine}
Robotics, M.: Milrem robotics to deliver 14 themis ugvs to ukraine in cooperation with kmw (2022), \url{https://milremrobotics.com/milrem-robotics-to-deliver-14-themis-ugvs-to-ukraine-in-cooperation-with-kmw/}

\bibitem{themis_ugv}
Robotics, M.: The themis ugv (2024), \url{https://milremrobotics.com/defence/}

\bibitem{salvato2021characterization}
Salvato, E., Fenu, G., Medvet, E., Pellegrino, F.A.: Characterization of modeling errors affecting performances of a robotics deep reinforcement learning controller in a sim-to-real transfer. In: 2021 44th International Convention on Information, Communication and Electronic Technology (MIPRO). pp. 1154--1159. IEEE (2021)

\bibitem{ieeRealityGap}
Salvato, E., Fenu, G., Medvet, E., Pellegrino, F.A.: Crossing the reality gap: A survey on sim-to-real transferability of robot controllers in reinforcement learning. IEEE Access  \textbf{9},  153171--153187 (2021)

\bibitem{stefanova2018off}
Stefanova, Z.S., Ramachandran, K.M.: Off-policy q-learning technique for intrusion response in network security. World Academy of Science, Engineering and Technology, International Science Index  \textbf{136},  262--268 (2018)

\bibitem{veranda_git}
Stelter, A., Diezel, K., Williams, S., Sutton, R., Smith, C.: Veranda: A 2-dimensional mobile robotics simulation environment (2018), \url{https://github.com/roboscienceorg/veranda}

\bibitem{sutton2018reinforcement}
Sutton, R.S., Barto, A.G.: Reinforcement learning: An introduction. MIT press (2018)

\bibitem{DevelopmentOfAutonomy}
Thoresen, M., Baksaas, M., Hanevold, M., Kolden, D., Mathiassen, K., Nielsen, N.H.: {Development of autonomy for off-road driving for unmanned ground vehicles}. {FFI}-{R}eport 24/00346, {Norwegian Defence Research Establishment} (TBP), ugradert

\bibitem{ugv_reward}
Vahtla, A.: Russia announces reward for capture of estonian-made milrem ugv in ukraine (2022), \url{https://news.err.ee/1608705544/russia-announces-reward-for-capture-of-estonian-made-milrem-ugv-in-ukraine}

\bibitem{zennaro2023modelling}
Zennaro, F.M., Erd{\H{o}}di, L.: Modelling penetration testing with reinforcement learning using capture-the-flag challenges: Trade-offs between model-free learning and a priori knowledge. IET Information Security  \textbf{17}(3),  441--457 (2023)

\end{thebibliography}

\end{document}